\documentstyle{article}
\newcommand{\hochpunkt}[1]{\mbox{$^{\raisebox{.3ex}{\scriptsize #1}}_{\raisebox{.6ex}{\hspace{.17em}.}}$}}
\newcommand{\hoch}[1]{\mbox{$^{\raisebox{.3ex}{\scriptsize #1}}$}}
\begin{document}

\begin{center}
{\LARGE \bf Photometry of some neglected bright cataclysmic variables and
candidates}\footnote{Based on observations taken at the Observat\'orio do 
Pico dos Dias / LNA}

\vspace{1cm}  

{\Large \bf Albert Bruch}

\vspace{0.5cm}
Laborat\'orio Nacional de Astrof\'{i}sica, 
Rua Estados Unidos, 154, CEP 37504-364, Itajub\'a - MG, Brazil

\vspace{0.8cm}

(Accepted for publication in New Astronomy)
\vspace{0.8cm}
\end{center}

\begin{abstract}
As part of an effort to better characterize bright cataclysmic variables 
(CVs) which have received little attention in the past light curves of four 
confirmed systems (CZ~Aql, BO~Cet, V380~Oph and EF~Tuc) and one candidate 
(Lib~3) are analyzed. For none of these stars time resolved photometry has 
been published previously. While no variability was found in the case of
Lib~3, which thus cannot be confirmed as a CV, the light curves of all other 
targets are dominated by strong flickering. Modulations on hourly time scales
superimposed on the flickering can probably be related to orbital variations
in BO~Cet and V380~Oph, but not in CZ~Aql and EF~Tuc. Variations on the time
scale of 10 minutes in CZ~Aql, while not yet constituting convincing evidence,
together with previous suspicions of a magnetically channeled accretion
flow may point at an intermediate polar nature of this star. Some properties
of the flickering are quantified in an effort to enlarge the data base for
future comparative flickering studies in CVs and to refine the classification
of the target stars. 
\vspace{1ex}

{\parindent0em {\bf Keywords:}
Stars: novae, cataclysmic variables --
Stars: individual: CZ~Aql -- 
Stars: individual: BO~Cet --
Stars: individual: V380~Oph -- 
Stars: individual: EF~Tuc -- 
Stars: individual: Lib~3}

\end{abstract}

\section{Introduction}
\label{Introduction}

Cataclysmic variables (CVs) are binary stars where a Roche-lobe filling 
late-type component (the secondary) transfers matter via an accretion disk 
to a white dwarf primary. It may be surprising that even after decades of 
intense studies of CVs there are still an appreciable number of known or 
suspected systems, bright enough to be easily observed with comparatively small
telescopes, which have not been studied sufficiently for basic parameters
such as the orbital period to be known with certainty. In some cases even 
their very class membership is not confirmed. 

Therefore, I started a small observing project aimed at a better understanding
of these so far neglected stars. First results have been published in 
Bruch (2016) and Bruch (2017). Here, I discuss light
curves of some further targets of the project for which no time resolved
photometry has yet been published. They were observed in order to verify 
the presence of flickering typically observed in CVs and to try to measure
the orbital period photometrically (or to confirm suspected periods). 

It is well known that accretion of mass via a disk onto a central object
normally leads to apparently stochastic brightness variations termed
flickering. It occurs to a more or less obvious degree in objects
as diverse as Active Galactic Nuclei (Garcia et al. 1999), certain stages of 
star formation (Herbst \& Sevchenko 1999, Kenyon et al. 2000, Scaringi et al
2015), x-ray binaries (van der Klis 2004) or some (but not all) symbiotic 
stars (Gromadzki et al. 2006).
The time scales of such variations and the spectral range in which
they most prominently appear depend on the nature of the particular system.
In the optical range flickering is by far most conspicuous in CVs where it
leads to variability typically on time scales of the order of minutes and 
with amplitudes which can range from a few millimagnitudes to more than an 
entire magnitude. For a general characterization of flickering in CVs, see 
Bruch (1992).

While the strength of the flickering, as measured by its total amplitude,
depends heavily on the individual system and its momentary photometric
state no photometric observations of well established CVs obtained with a 
suitable time resolution and signal-to-noise ratio, of which the present
author is aware, ever showed the absence of flickering, unless the
respective system was in a state of (temporary) suspension of mass accretion
onto the white dwarf. Thus, the presence of flickering appears to be a 
necessary (but not sufficient) condition for a star to be classified as 
a cataclysmic variable.

While flickering by itself is a fascinating and still not well understood
phenomenon, it can also be a severe obstacle to find and characterize modes of
variability in CVs which may have similar amplitudes but are due to different 
mechanisms. Thus, orbital variations caused by the changing aspect of the
system around the orbit may easily be masked by flickering. This is the
case in particular in systems with a low or intermediate orbital inclination
where such variations remain small and in the presence of flickering may only
be detected when observations over many cycles are averaged. 

The objects discussed in the present paper are four systems taken from
the most recent on-line version of the Ritter \& Kolb catalogue
\cite{Ritter03} which have only unconfirmed or uncertain orbital periods. 
These are CZ~Aql, BO~Cet, V380~Oph and 
EF~Tuc. To these I add Lib~3 (= Preston 874124),
an unconfirmed CV listed in the catalogue of Downes et al. 2005 and 
classified as being of UX~UMa subtype.

In Sect.~\ref{Observations and data reductions} the observations and data
reduction techniques are briefly presented. Sects.~\ref{CZ Aql} -- \ref{Lib 3}
deal with the individual objects of this study, focusing on orbital 
variations, while in 
Sect.~\ref{Variations on short time scales} the rapid variations observed in
four of the five targets are collectively quantified. Finally, a short summary
in Sect.~\ref{Summary} concludes this paper.

\section{Observations and data reductions}
\label{Observations and data reductions}

All observations were obtained at the 0.6-m Zeiss and the 0.6-m Boller \& 
Chivens telescopes of the Observat\'orio do Pico dos Dias, operated by 
the Laborat\'orio Nacional de Astrof\'{\i}sica, Brazil. 
Time series imaging of the field around the target stars was performed
using cameras of type Andor iKon-L936-B and iKon-L936-EX2 equipped with 
back illuminated, visually optimized CCDs.
In order to resolve the expected rapid flickering variations the integration
times were kept short. Together with the small readout times of the detectors
this resulted in a time resolution of the order of 5\hoch{s}. In order to 
maximize the count rates in these short time intervals no filters were
used. Therefore, it was not possible to calibrate the stellar magnitudes.
Instead, the brightness is expressed as the magnitude difference between the 
target and a nearby comparison star. This is not a severe limitation in view 
of the purpose to the observations. A rough estimate of the effective 
wavelength of the white light band pass, assuming a mean atmospheric extinction 
curve, a flat transmission curve for the telescope, and a detector efficiency 
curve as provided by the manufacturer, yields $\lambda_{\rm eff} \approx
5530 \AA$, very close to the effective wavelength of the Johnson $V$ band
(5500~\AA; Allen 1973).

A summary of the observations is 
given in Table~\ref{Journal of observations}. Some light curves contain
gaps caused by intermittent clouds or technical reasons. Basic data 
reduction (biasing, flat-fielding) was performed using IRAF. 
For the construction of light curves aperture photometry routines 
implemented in the MIRA software system (Bruch 1993) were employed. The
same system was used for all further data reductions and calculations. 
Throughout this
paper time is expressed in UT. However, whenever observations taken in 
different nights were combined (e.g., to search for orbital variations)
time was transformed into barycentric Julian Date on the Barycentric
Dynamical Time (TDB) scale using the online tool provided by 
Eastman et al.\ (2010) in order to take into account variations of the 
light travel time within the solar system. Timing analysis of the data
employing Fourier techniques was done using the Lomb-Scargle 
algorithm (Lomb 1976, Scargle 1982, Horne \& Baliunas 1986) unless specified
otherwise. The terms ``power spectrum'' and ``Lomb-Scargle periodogram'' are
used synonymously for the resulting graphs.

\begin{table}

\caption{Journal of observations}
\label{Journal of observations}

\hspace{1ex}

\begin{tabular}{llcc}
\hline
Name      & Date & Start & End  \\
          &      & (UT)  & (UT) \\
\hline
CZ Aql    & 2014 Jun 17 & \phantom{2}3:31 & \phantom{2}8:54 \\
          & 2014 Jun 19 & \phantom{2}2:40 & \phantom{2}8:31 \\
          & 2014 Aug 25/26 &        22:46 & \phantom{2}4:06 \\
          & 2014 Sep 21/22 &        22:16 & \phantom{2}2:28 \\
          & 2014 Sep 22 &           21:44 & 23:12 \\
          & 2014 Sep 23 &           21:47 & 23:13 \\
          & 2014 Sep 24 &           21:56 & 23:22 \\ [1ex]
BO Cet    & 2014 Sep 23 & \phantom{2}5:21 & \phantom{2}8:14 \\
          & 2014 Sep 24 & \phantom{2}5:12 & \phantom{2}8:10 \\
          & 2014 Oct 23 & \phantom{2}1:31 & \phantom{2}3:10 \\
          & 2016 Aug 11 & \phantom{2}5:49 & \phantom{2}8:51 \\
          & 2016 Aug 12 & \phantom{2}5:55 & \phantom{2}8:44 \\ [1ex]
Lib 3     & 2016 Jun 27/28 & 21:16        & 2:55 \\ [1ex]
V380 Oph  & 2014 Jun 18 & \phantom{2}1:00 & \phantom{2}6:37 \\
          & 2014 Jun 20 & \phantom{2}4:36 & \phantom{2}6:05 \\
          & 2014 Jun 22 & \phantom{2}2:37 & \phantom{2}4:30 \\ [1ex]
EF Tuc    & 2014 Jun 18 & \phantom{2}8:02 & \phantom{2}8:44 \\
          & 2014 Jun 23 & \phantom{2}4:06 & \phantom{2}8:56 \\
          & 2014 Aug 26 & \phantom{2}4:50 & \phantom{2}8:43 \\
          & 2014 Sep 22 & \phantom{2}2:39 & \phantom{2}8:13 \\
          & 2014 Sep 22/23 &        23:20 & \phantom{2}5:11 \\
          & 2014 Sep 23/24 &        23:22 & \phantom{2}5:05 \\
          & 2014 Sep 24/25 &        23:27 & \phantom{2}2:19 \\
          & 2015 Aug 11 & \phantom{2}7:40 & \phantom{2}8:47 \\
          & 2015 Aug 12 & \phantom{2}7:18 & \phantom{2}8:41 \\
          & 2015 Aug 13 & \phantom{2}6:52 & \phantom{2}8:51 \\
          & 2015 Aug 14 & \phantom{2}7:06 & \phantom{2}8:59 \\
          & 2015 Aug 15 & \phantom{2}7:01 & \phantom{2}8:41 \\ [1ex]
\hline
\end{tabular}
\end{table}
%

\section{CZ Aql}
\label{CZ Aql}

CZ~Aql was discovered as a variable star by Reinmuth (1925) who did 
not provide a classification. Based on spectroscopic evidence, 
Cieslinski et al.\ (1998) considered the star to be a dwarf nova and 
suspected an orbital period of 4.8 hours. This is confirmed by 
Sheets et al.\ (2007) who find 
$P_{\rm orb} = 0.2005$ days = 4.812 hours but cannot distinguish 
between aliases ranging between 4.798 and 4.826 hours. Based on a detailed
spectroscopic analysis they suspect CZ~Aql to contain a magnetically channeled 
accretion flow. 

\begin{figure}
   \parbox[]{0.1cm}{\epsfxsize=14cm\epsfbox{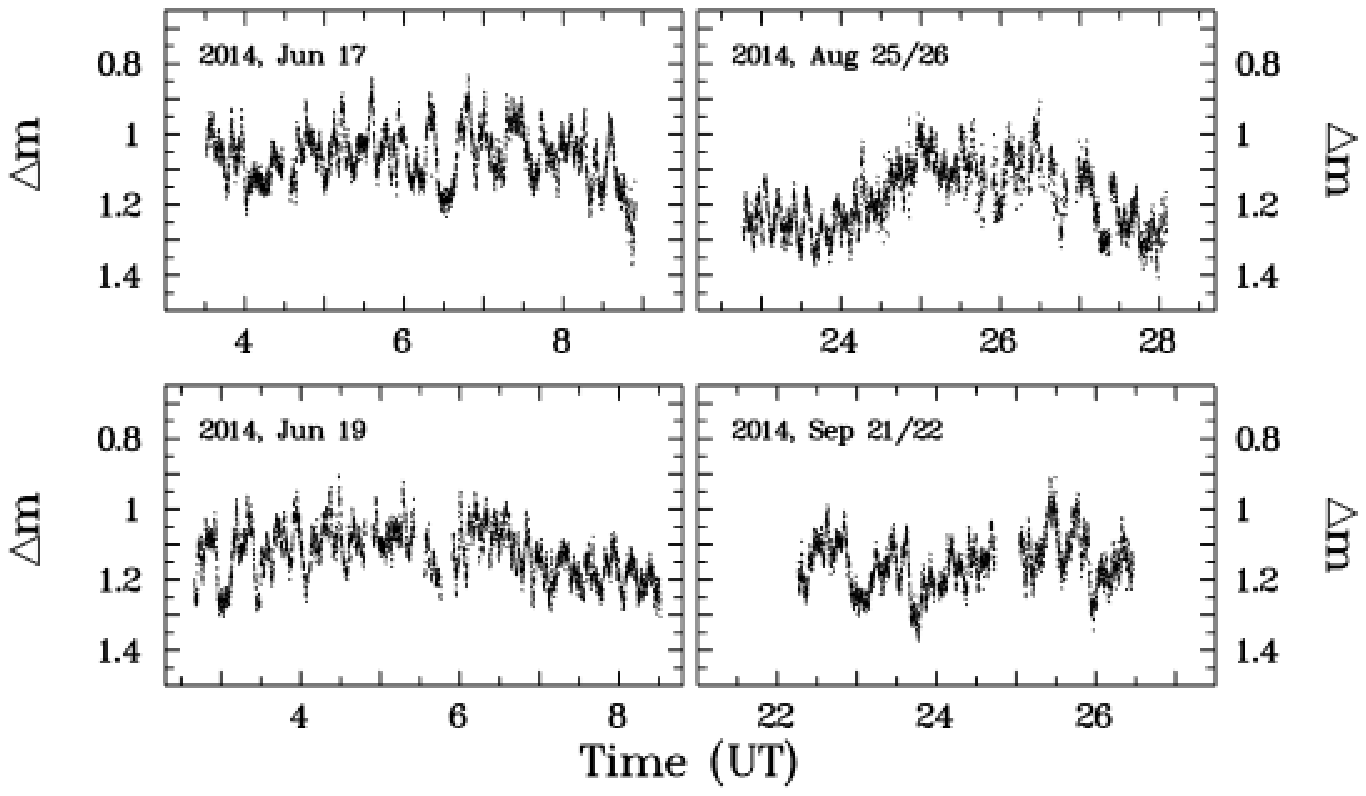}}
      \caption[]{Light curves of CZ Aql 
                 observed in four nights in 2014.}
\label{czaql-lightc}
\end{figure}

CZ~Aql was observed in seven nights between 2014, June 17 and September 24. 
Representative light curves, drawn on the same time and magnitude scale to
ease comparison, are shown in Fig.~\ref{czaql-lightc}. 
Differential magnitudes are given with respect to the comparison star 
UCAC4 415-122382 ($V=14\hochpunkt{m}122$; Zacharias et al. 2013).
This translates into a rough average nightly
magnitude of CZ~Aql between $15\hochpunkt{m}0$ and $15\hochpunkt{m}3$
which is on the faint end of the distribution of the few magnitude estimates
in the data base of the American Association of Variable Star Observers 
(AAVSO). 

\input epsf

The light curves are dominated by flickering superposed on more 
gradual variations on the time scale of several hours, most clearly visible 
on 2014, Aug. 25 (Fig.~\ref{czaql-lightc}). In fact, adjusting a sine curve 
to the data of that night alone gives a best fit at a
period of 4.704 hours, just slightly shorter than but still compatible with
the spectroscopic period determined by Sheets et al.\ (2007).
In order to investigate if the variations can be explained as a periodic 
modulation all light curves were combined into a single data set after 
subtracting the mean differential magnitude
of the individual nights in order to remove long term brightness changes.
The final part of the light curve of 2014 June 17 (Fig.~\ref{czaql-lightc})
was not considered since the sudden drop observed at the very end may in part 
be due to colour differences between
CZ~Aql and the comparison star, the observations having been performed at an
air mass $>2$. 

\begin{figure}
   \parbox[]{0.1cm}{\epsfxsize=14cm\epsfbox{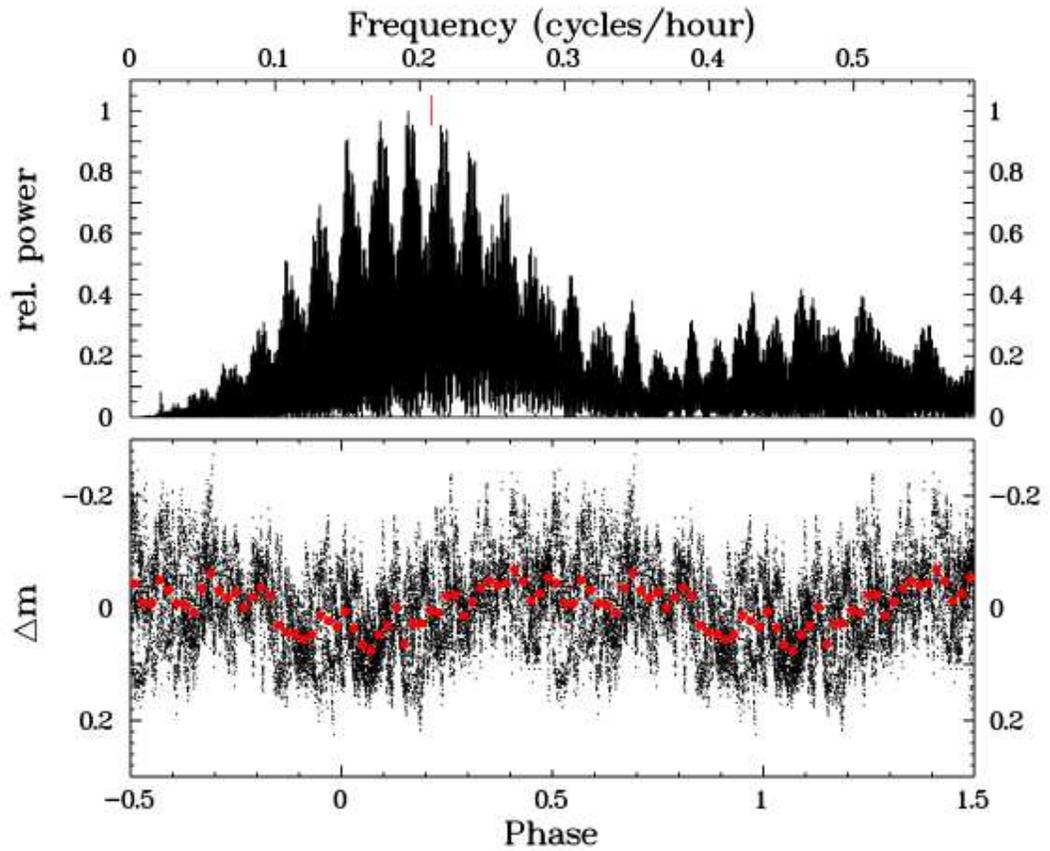}}
      \caption[]{{\it Top:} Power spectrum of the combined light curves of 
                 CZ~Aql of 2014, June 17 and 19. The red vertical line marks
                 the spectroscopic period. {\it Bottom:} Light curves 
                 folded on the period $P = 5.2083$ hours, corresponding
                 to the highest peak in the power spectrum. The red dots
                 represent a binned version of the folded light curve.}
\label{czaql-pow-fold}
\end{figure}

A power spectrum of the combined light curve was calculated (upper frame of 
Fig.~\ref{czaql-pow-fold}). Due to the distribution of the light curves along 
three months, it contains strong alias patterns with a broad maximum roughly
consistent with the spectroscopic period. However, none of the alias peaks
coincides with that period (marked by the red vertical line in the figure)
or lies within the possible range quoted by
Sheets et al.\ (2007). Instead, the frequency of the strongest peak 
corresponds to a period of 5.2083 hours\footnote{I do not claim that this is
the ``correct'' alias choice. Other alias peaks, corresponding to periods 
both, longer and shorter than the spectroscopic period, are almost as strong.},
$\sim$8\% longer than the orbital period. 
Folding the combined light curves on this period (lower frame of 
Fig.~\ref{czaql-pow-fold}) results in a consistent, approximately sinusoidal
pattern, modulated on small scales in phase by the strong flickering of the
star. But note that the light curve folded on periods corresponding to other
peaks in the power spectrum exhibit very similar patterns.

While the folded light curve thus suggests consistent variations of CZ~Aql
over a time scale of several months, it would be premature to 
consider them firmly established. Similar variations with periods not
coinciding with the orbital period have been observed in other cataclysmic
variables and have shown not be to stable on longer time scales. Examples
include 
V603~Aql (Haefner \& Metz 1985, Bruch 1991, Patterson et al.\ 1993),
TT~Ari (Belova et al.\ 2013 and references therein; Smak 2013),
KR~Aur (Kozhevnikov 2007),
V751~Cyg (Patterson et al.\ 2001, Papadaki et al.\ 2009),
V795~Her (Papadaki et al.\ 2006), and
V378~Peg (Kozhevnikov 2012).
Such variations are often interpreted as positive or negative superhumps. 
Long light curves observed during subsequent nights should permit to
verify if the variability of CZ~Aql follows similar patterns.

\section{BO Cet}
\label{BO Cet}
 
BO~Cet is first mentioned as a variable star and classified as a novalike
variable in the 71th name list of variable stars (Kazarovetz et al. 1993) with
a reference to a manuscript by R.\ Remillard. Similarly, in the on-line edition
of their catalog and atlas of cataclysmic variables Downes et al.\ (2005) cite
a private communication of R.\ Remillard as type reference for the star.

There is no doubt about the nature of BO~Cet as a cataclysmic variable.
The colours measured by Zwitter \& Munari (1995) ($U-B = -0.7$; $B-V = 0.1$;
$V-R = 0.07$; $R-I = 0.10$) are quite normal for a CV where the contribution
of the secondary star is negligible. The same authors also reproduce a 
spectrum of the star
which exhibits a strong blue continuum superposed by hydrogen and He~I emission 
lines. Moreover, a rather strong He~II $\lambda$~4686~\AA\, emission is also
present, reaching an integrated flux similar to that of H$\delta$. Based on
time resolved spectroscopic observations Rodr\'{\i}guez-Gil et al.\ (2007)
classified BO~Cen as a SW~Sextantis type star. The orbital period of 
0.13980 $\pm$ 0.00006 days (3\hochpunkt{h}3552) is only known from an 
informal communication by J.\, 
Patterson\footnote{http://cbastro.org/communications/news/messages/0274.html; 
or http://cbastro.org/pipermail/cba-public/2002-October/000300.html} based on
Center for Backyard Astrophysics (CBA) data.

\begin{figure}
   \parbox[]{0.1cm}{\epsfxsize=14cm\epsfbox{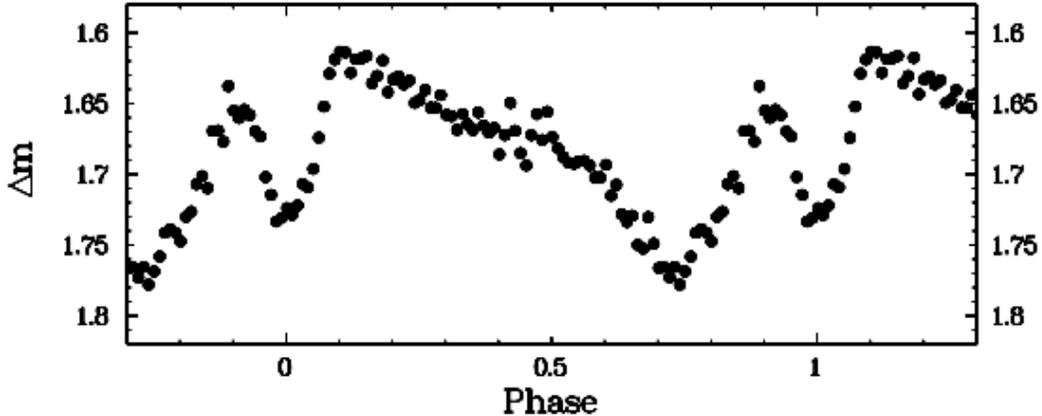}}
      \caption[]{Average orbital light curve of BO~Cet based on CBA data.}
\label{bocet-cba}
\end{figure}

Since the CBA observations are not otherwise documented I retrieved the light 
curve of BO~Cet from Patterson's communication. Phase Dispersion Minimization
(Stellingwerf 1978) as well as Analysis of Variance 
(Schwarzenberg-Czerny 1989),
of these data, both better suited for period searches in non-sinusoidally 
varying signals than Fourier transforms, yield a slightly longer period (but
still comfortably within the error limits) of 
$P_{\rm orb} = 0.13983$ days than
that informed by Patterson. The difference implies a phase shift of 0.05
over the time base of the CBA observations. Therefore, the phase folded
light curve, reproduced in Fig.~\ref{bocet-cba} is slightly different (in
fact, ``nicer'', particularly at the primary minimum) than the one based on
Patterson's period. If the phase of the secondary minimum, interpreted by 
Patterson as a shallow eclipse, is defined as phase $\phi = 0$, the average 
light curve assumes a minimum (deeper than the eclipse) at $\phi = 0.75$, 
rises steeply until $\phi = 0.1$ (this rise being interrupted by the eclipse)
and then declines first slowly and then more rapidly after $\phi = 0.5$. 
This waveform is different from the orbital variations of most CVs which 
often (but not always) exhibit a hump caused by a hot spot in the range 
$0.75 \le \phi \le 1$.

\begin{figure}
   \parbox[]{0.1cm}{\epsfxsize=14cm\epsfbox{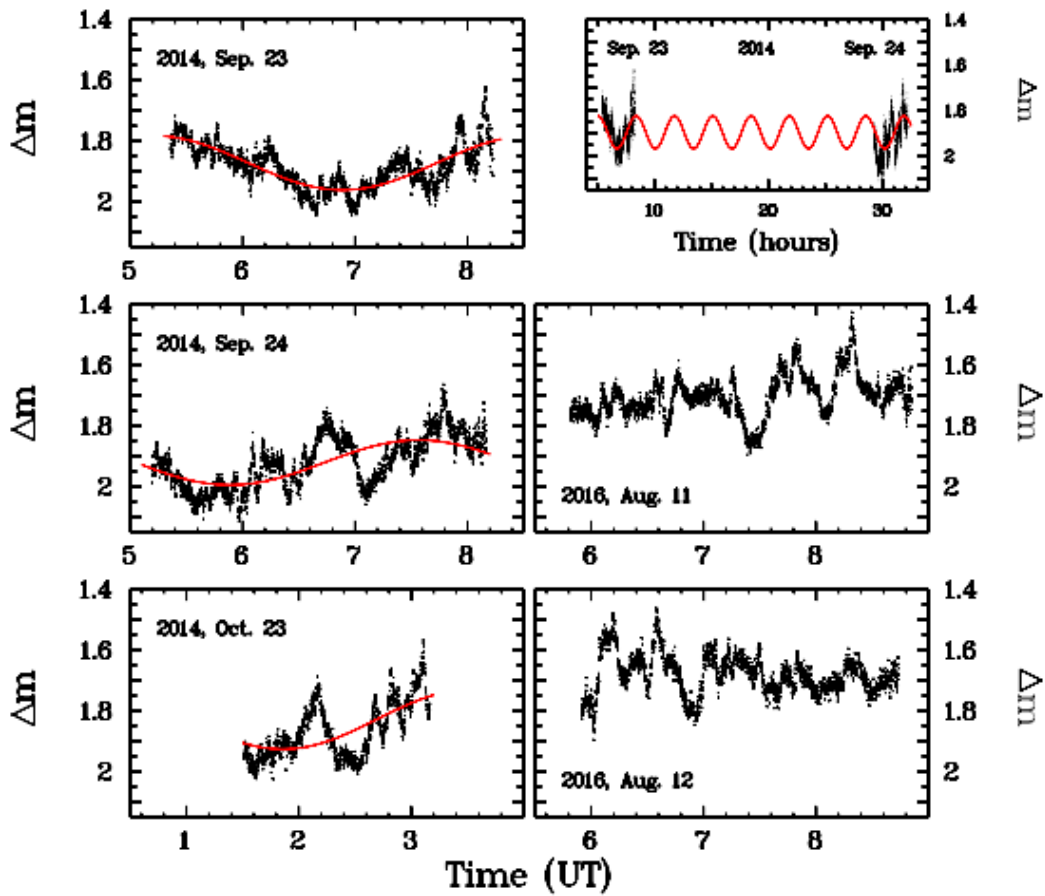}}
      \caption[]{Light curves of BO~Cet observed in 2014 (frames on the
                 left side) and 2016 (lower two frame on the right side).
                 The red curves are sine fits to the data, where the
                 period has been fixed to $P_{\rm orb}$. The 
                 upper right frame contains the combined light curves of 2015, 
                 Sep. 24 and 25 together with the best fit sine curve, 
                 showing that the adopted period does not lead to a significant
                 phase shift between the data of the two nights.}
\label{bocet-lightc}
\end{figure}

The absence of published time resolved photometry of BO~Cet prompted me to 
include the star in the present observing project. I observed it in five
nights in 2014 and 2016 (see Table~\ref{Journal of observations}),
obtaining light curves with durations of the order of 3\hoch{h} (1\hoch{h}
in one case) which is close to the reported orbital period. They are shown
(all drawn on the same time and magnitude scale in order to ease comparison) 
in Fig.~\ref{bocet-lightc}. Differential magnitudes are given with respect 
to the comparison star UCAC4 441-002697 
($V=12\hochpunkt{m}720$; Zacharias et al.\ 2013). The average nightly
magnitudes of BO~Cet thus varied roughly between $14\hochpunkt{m}5$ and 
$14\hochpunkt{m}7$, well within the brightness range quoted in the data base
of the the Association Fran\c{c}aise des Observateurs d'Etoiles Variables 
(AFOEV; 13\hochpunkt{m}6 -- 15\hochpunkt{m}2) and the British Astronomical
Association, Variable Star Section (BAAVSS; 14\hochpunkt{m}3 -- 
14\hochpunkt{m}). 

BO~Cet is characterized by significant flickering
as is normal for cataclysmic variables. Moreover, the light curves observed
in 2014 exhibit more gradual variations which are compatible with a modulation
on the orbital period. This is highlighted in the figure by the red
curves which are best sine fits to the data (as a rough approximation to the 
orbital variations) where the period has been fixed
to $P_{\rm orb}$. The amplitude of this modulation, measured to be 
0\hochpunkt{m}089 (0\hochpunkt{m}074; 0\hochpunkt{m}095) on 2016, Sep.\ 23 
(Sep. 24, Oct. 23), may be considered constant regarding the uncertainties 
expected in view of the flickering activity. 

The upper right frame of Fig.~\ref{bocet-lightc} shows the combined light
curves observed on 2014, Sep. 23 and 24 together with the best fit sine
curve, again with the period fixed to $P_{\rm orb}$. While the variations on
Sep. 24 are not perfectly in phase with those of the previous night the
difference is not large enough to suggest a significant period error, again
considering the strong flickering.

Compared to 2014, in the 2016 observations the orbital modulation is not as 
well expressed. The overall magnitude level being slightly higher may not be
significant since different detectors were used during the two epochs which
introduces uncertainties in the differential magnitudes of these uncalibrated
white light measurements.

\section{V380 Oph}
\label{V380 Oph}

V380~Oph was identified as a variable star by Hoffmeister (1929). Its 
spectrum is considered as a ``textbook example'' for a CV by 
Liu et al.\ (1999b)\footnote{Is is not quite clear who first classified the 
star as a CV.}. X-rays from the source were detected in the ROSAT All Sky 
Survey (Verbunt et al. 1997).

The orbital period was spectroscopically determined by Shafter (1983, 1985) to
be 0.16 days; a value later refined to 0.154107 days (3.6986 hours) by 
Rodr\'{\i}guez-Gil et al.\ (2007) who also classified V380~Oph as a SW~Sex 
star. A long-term light curve with rather smooth variations between 
$\sim$$14\hochpunkt{m}3$ and $16\hoch{m}$ is shown by Kafka \& Honeycutt 
(2004), but an excursion to a low state of $\sim$$17\hochpunkt{m}5$ observed by 
Shugarov et al.\ (2005) and in particular light curves generated from more 
recent data provided by the AAVSO, the AFOEV and the BAAVSS also justify to 
count V380~Oph among the VY~Scl stars.

\begin{figure}
   \parbox[]{0.1cm}{\epsfxsize=14cm\epsfbox{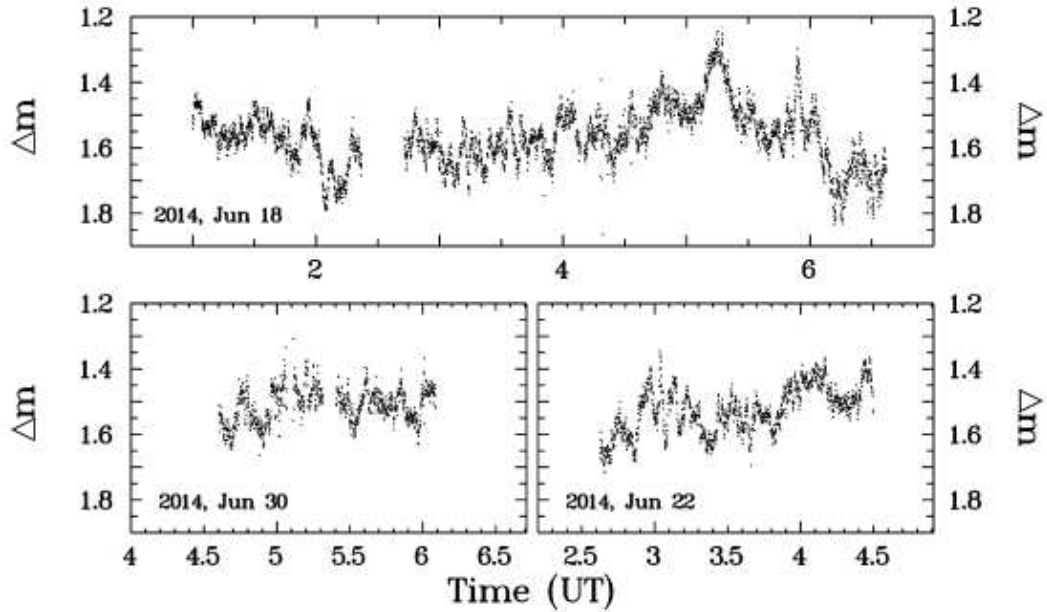}}
      \caption[]{Light curves of V380~Oph during
                 three nights in 2014 June.}
\label{v380oph-lightc}
\end{figure}

V380~Oph was observed in the three nights of 2014 June 18, 20 and 22. The
light curves, expressed as differential magnitudes with respect to the primary
comparison star UCAC4 481-069940 are shown in Fig.~\ref{v380oph-lightc}.
This star having a magnitude of $V=13\hochpunkt{m}911$ (Zacharias et al. 2013), 
a mean brightness of V380~Oph of roughly 15\hochpunkt{m}4 with only
slight night-to-night variations is calculated. This is slightly fainter 
than the average brightness of the system according to 
Kafka \& Honeycutt (2004) and the long term light curves found in the AAVSO,
AFOEV and BAAVSS data bases (not considering low states).

\begin{figure}
   \parbox[]{0.1cm}{\epsfxsize=14cm\epsfbox{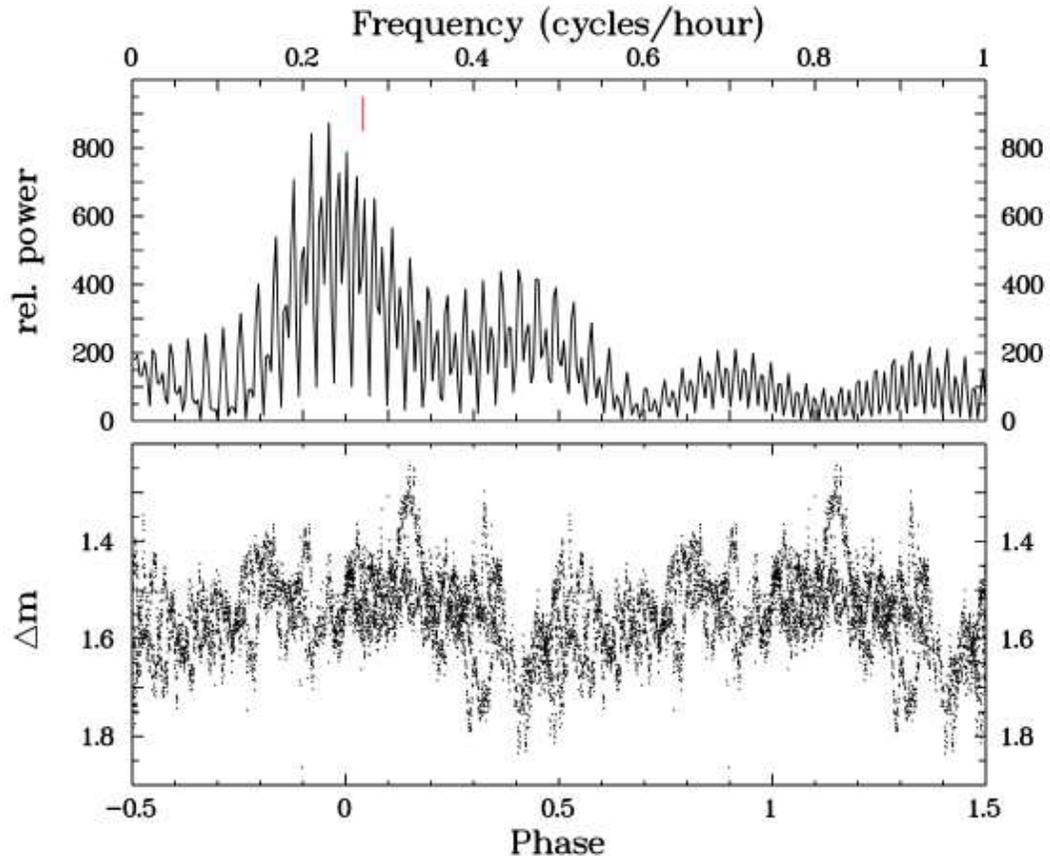}}
      \caption[]{{\em Top:} Power spectrum of the combined light
                 curves of V380~Oph on 2014, June 18, 20 and 22. The
                 red vertical line marks the orbital frequency. 
                 {\em Bottom:} Combined light curve folded on the 
                 orbital period.}
\label{v380oph-fold}
\end{figure}

The light curves are characterized by significant flickering activity,
superposed on variations on hourly time scales. In order to investigate
if the latter are compatible with the orbital period, the three light
curves were combined into a single data set (without subtracting the
average nightly magnitude) and subjected to a Fourier analysis. The
low frequency part of the power spectrum is shown in the upper frame
of Fig.~\ref{v380oph-fold}. Overlaid upon the expected alias pattern due to the
separation of two nights between the individual light curve segments is a 
low frequency signal which is roughly compatible with the orbital period
determined by Rodr\'{\i}guez-Gil et al.\ (2007). The frequency corresponding
to that period is indicated by a broken vertical line in the figure. While
it does not coincide with the highest alias peak it is well aligned with a
secondary maximum\footnote{In view of the strong flickering and the limited 
number of photometric data the fact that the highest peak is not at the
orbital frequency is no indication that the spectroscopic period of 
Rodr\'{\i}guez-Gil et al.\ (2007) is in error. However, considering the
alias pattern in their Fig.~12 and the distribution of the observations
as deduced from their Table 1 it cannot be excluded that the true period
of V380~Oph corresponds to an alias of the power spectrum peak chosen by
them.}. Folding the combined light curves on the orbital period (lower
frame of Fig.~\ref{v380oph-fold}) suggests that the brightness variations on
hourly time scales have indeed an orbital origin. 

\section{EF Tuc}
\label{EF Tuc}

EF~Tuc was discovered as a dwarf nova in the Edinburgh-Cape Blue Object 
Survey. The spectrum shown by Stobie et al.\ (1995) is dominated by strong
hydrogen emission lines. Chen et al.\ (2001) investigated the spectrum in
somewhat more detail and suspected a SU~UMa type classification. They could
not measure the orbital period spectroscopically, probably because of a low
orbital inclination. The orbital period of 3.6 hours 
quoted in the Ritter \& Kolb 
catalogue is based on informal communications by J.\ Patterson in 2003
and 2006\footnote{http://cbastro.org/communications/news/messages/0350.html;
http://cbastro.org/communications/news/messages/0487.html} where it is
classified as a candidate period. 
It has never been confirmed. It would contradict the SU~UMa classification
because stars of that type have periods below the 2.2 -- 2.8 hour period gap in
the CV orbital period distribution. Chen et al.\ (2001) also mention time
resolved photometry and found that EF~Tuc exhibits strong flickering, but
they do not show a light curve.

\begin{figure}
   \parbox[]{0.1cm}{\epsfxsize=14cm\epsfbox{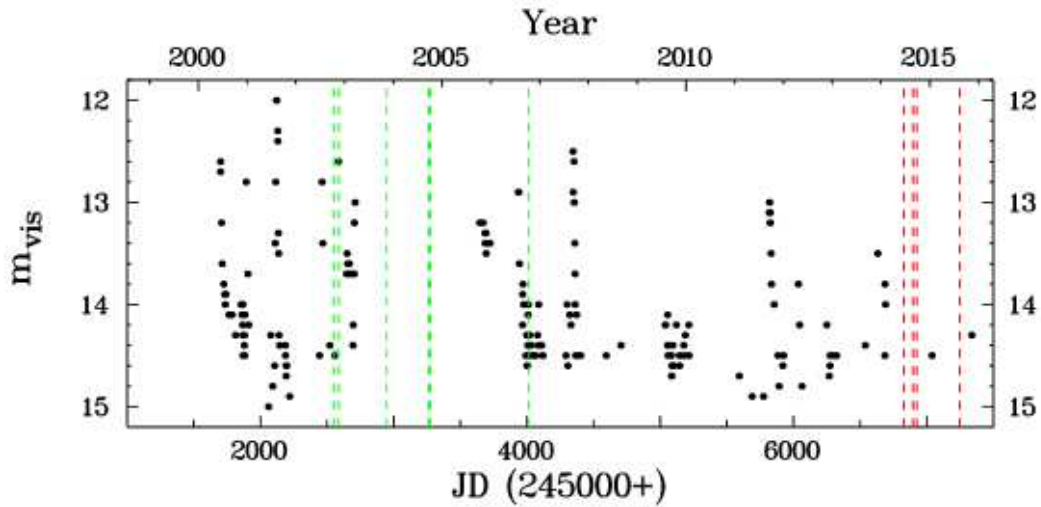}}
      \caption[]{AAVSO long term light curve of EF Tuc. The broken
                 vertical lines indicate the epochs of AAVSO high cadence
                 observations (green) and of the light curves listed in
                 Tab.~\ref{Journal of observations} (red).}
\label{eftuc-aavso}
\end{figure}

The AAVSO long term light curve, shown in Fig.~\ref{eftuc-aavso}, although not
having a dense temporal coverage, exhibits a the typical dwarf nova
behaviour with outbursts reaching $\sim$$12\hoch{m} - 12\hochpunkt{m}5$ 
and a quiescent level between 14\hochpunkt{m}5 and 15\hoch{m}.
While the coverage is insufficient
to verify if some of the outbursts discernable in the figure are
superoutbursts or not, their amplitudes are certainly quite low for such
an interpretation. The outburst with the best coverage, peaking close to
JD~2452120, has a total width of $\sim$40 days which is long even for a
superoutburst of an SU~UMa star (see, e.g., the 
compilation of Kato et al. 2009). Moreover, unlike observed
in superoutbursts, the rise to maximum takes more than 10 days, and there is 
no plateau phase after maximum. Thus the long term light curve cannot sustain 
the SU~UMa classification of EF~Tuc.

\begin{figure}
   \parbox[]{0.1cm}{\epsfxsize=14cm\epsfbox{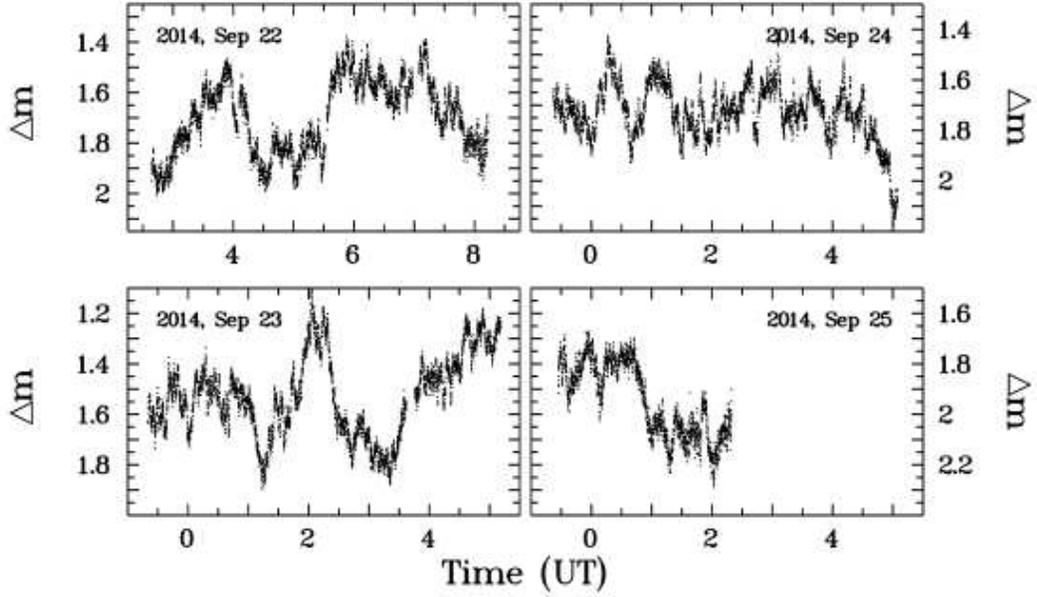}}
      \caption[]{Light curves of EF~Tuc in four nights in 2014, September.}
\label{eftuc-lightc}
\end{figure}

EF~Tuc was observed in 12 nights in 2014 and 2015. The respective epochs
are marked by red vertical lines in Fig.~\ref{eftuc-aavso}. The best data 
consist of
light curves obtained in four consecutive nights in 2014, September. They are
shown in Fig.~\ref{eftuc-lightc} on the same time and magnitude scale. 
Differential magnitudes are given with respect to the primary comparison star 
UCAC4 115-000029 ($V=13\hochpunkt{m}328$; Zacharias et al. 2013). The range 
of average nightly magnitudes of EF~Tuc is then 14\hochpunkt{m}5 -- 
15\hochpunkt{m}4. This is at the faint end of the magnitude distribution 
in the AAVSO long-term light curves (Fig.~\ref{eftuc-aavso}).

\begin{figure}
   \parbox[]{0.1cm}{\epsfxsize=14cm\epsfbox{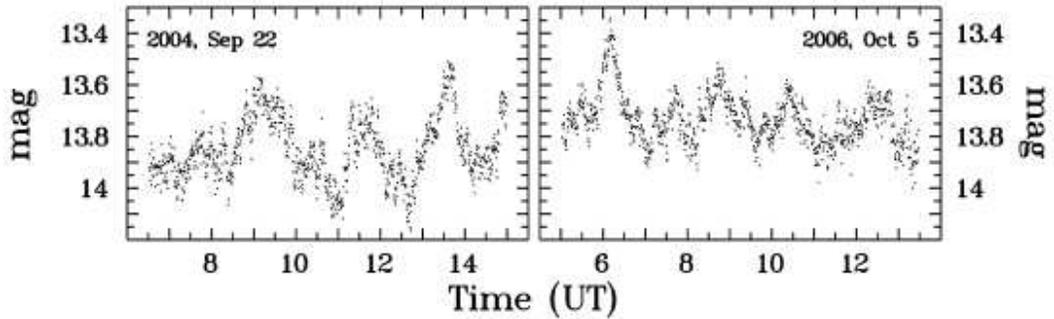}}
      \caption[]{AAVSO high cadence light curves of EF Tuc of 2005, Sep 22
                 and 2006, Oct. 5.}
\label{eftuc-aavso-hc}
\end{figure}

The light curves are dominated by strong flickering which is superposed
on variations occurring on hourly time scales. The total amplitude observed
in a single light curve reaches 1\hochpunkt{m}0. On average it is 
0\hochpunkt{m}63. Strong variations can occur quite rapidly: On 2014, Jul 23
a drop in magnitude of 0\hochpunkt{m}7 was observed within 35 minutes (see
also the strong and rapid flare on 2014, Sep.\ 23; lower left frame of
Fig.~\ref{eftuc-lightc}).

Apart from the long term light curve shown in Fig.~\ref{eftuc-aavso} the
AAVSO archives also contain several high cadence observations (observer:
Berto Monard). Although they
have a lower time resolution than the other EF~Tuc data discussed here they
contain complementary information. Therefore, I include them in the present 
study in particular to perform a search for periodic signals. Details of these 
light curves are given in Tab.~\ref{eftuc-aavso-tab}. Their epochs are marked
by green vertical lines in Fig.~\ref{eftuc-aavso}. Fig.~\ref{eftuc-aavso-hc} 
shows two representative examples.

\begin{table}

\caption{High cadence light curves of EF~Tuc in the AAVSO data base}
\label{eftuc-aavso-tab}

\hspace{1ex}

\begin{tabular}{lcccc}
\hline
Observing & Start & End    & Time      & Number     \\
Date      & (UT)  & (UT)   & Res. (s)  & of Integr. \\
\hline
2002 Oct 21 & \phantom{1}6:12 &           11:15 &  26 & \phantom{1}590 \\   
2002 Nov 08 & \phantom{1}5:17 &           12:37 &  16 &           1507 \\
2003 Oct 29 & \phantom{1}5:10 &           14:02 &  60 & \phantom{1}259 \\
2003 Nov 07 & \phantom{1}5:22 &           12:46 &  60 & \phantom{1}449 \\
2004 Sep 09 & \phantom{1}7:05 &           14:45 &  30 & \phantom{1}906 \\
2004 Sep 10 & \phantom{1}9:08 &           15:22 &  31 & \phantom{1}402 \\
2004 Sep 15 & \phantom{1}9:20 &           15:15 &  31 & \phantom{1}696 \\
2004 Sep 16 & \phantom{1}7:21 &           15:05 &  30 & \phantom{1}880 \\
2004 Sep 17 & \phantom{1}5:42 &           11:09 &  30 & \phantom{1}629 \\
2004 Sep 19 & \phantom{1}8:52 &           14:39 &  31 & \phantom{1}560 \\
2004 Sep 20 &           11:20 &           14:57 &  31 & \phantom{1}429 \\
2004 Sep 22 & \phantom{1}6:31 &           14:59 &  31 & \phantom{1}999 \\
2004 Sep 26 &           10:52 &           15:05 &  31 & \phantom{1}495 \\
2006 Oct 04 & \phantom{1}6:00 &           12:51 &  30 & \phantom{1}810 \\
2006 Oct 05 & \phantom{1}5:02 &           13:28 &  30 & \phantom{1}999 \\
2006 Oct 06 & \phantom{1}5:24 & \phantom{1}6:26 &  30 & \phantom{1}124 \\
\hline
\end{tabular}
\end{table}
%

In order to search for periodicities on hourly time scales, in  particular
in order to identify the orbital period of EF~Tuc, all light curves were 
subjected to a power spectrum analysis.
Additionally, power spectra of the combined light curves observed within
restricted periods of a couple of nights (such as those of 2004, Sep. 9 -- 26
or 2015, Sep. 22 -- 25) and subsets thereof were calculated. In these cases
night-to-night variations of the average magnitude were first subtracted.

The individual light curves are too short to reveal clear
periodicities on the time scale of hours; i.e.\ the range in which the
orbital period may be suspected. This is the more so in the presence of
strong flickering. However, it is intriguing that most individual power
spectra show peaks in the range between 0.35 and 0.45 cycles/hour.
In the case of the AAVSO light curves, a second peak centred on 0.15 
cycles/hour is often observed, which is missing in the power spectra of the
2014 -- 2015 data. The peaks are quite broad due to the limited frequency 
resolution caused by the finite lengths of the data trains. The power
spectra of the combined light curves repeat this pattern, but the peaks
are now split up into a many narrow aliases introduced by the complicated 
window function. 

\begin{figure}
   \parbox[]{0.1cm}{\epsfxsize=14cm\epsfbox{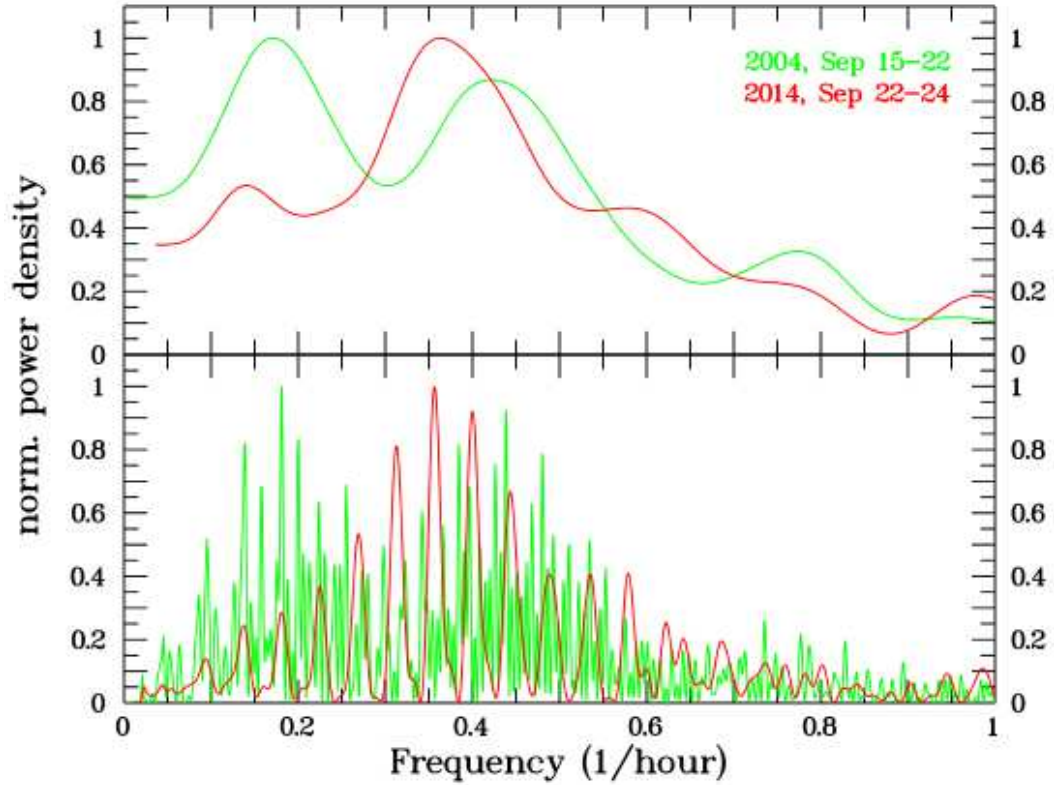}}
      \caption[]{Power spectra of the combined light curves of TF~Tuc of
                 2004, Sep 15-22 (green) and of 2015, Sep. 22-24 (red).
                 For details, see text.}
\label{eftuc-power}
\end{figure}

As examples, the summed individual power spectra\footnote{The power spectra
were weighted with the factor $\Delta t/n$, where $\Delta t$ is the entire
time base of the light curve and $n$ is the number of data points. Considering
that the Lomb-Scargle periodogram scales with $n$, this weighting scheme
ensures that each power spectrum gets a weight which increases with the
length of the light curve but is independent of the time resolution.}
of the nights of 2004, Sep. 15 -- 22 (green) and of 2014, Sep 22 -- 24
(red) are shown in the upper frame of Fig.~\ref{eftuc-power}. The lower
frame of the figure contains the power spectra of the combined light curves
in the respective time intervals.
Reflecting the trends already seen in the nightly
power spectra, in 2004 two almost equally strong clusters of peaks are
seen, the highest of which correspond to periods of 5\hochpunkt{h}52 and
2\hochpunkt{h}28, respectively. The low frequency cluster of peaks is
absent in the power spectrum of 2014. The much simpler alias pattern,
reflecting the smaller number of contributing nightly light curves (resulting
in a less complicated window function) is roughly consistent with the higher
frequency cluster observed in 2004, but appears to be shifted towards lower
frequencies and the individual peaks do not coincide with those of 2004. 
The highest one in 2015 corresponds to a period of 2\hochpunkt{h}80.

Folding the data on any of the periods does not result in a 
convincing light curve. The power spectra peaks apparently reflect
the presence of decimagnitude variations on the time scales of hours
such as those seen on 2014, Sep.\ 22 and 23 but which are not repeated 
in other nights such as 2014, Sep.\ 24 (see Fig.~\ref{eftuc-lightc}). It
may therefore be stated that these variations occur on a preferred time scale,
but they are not periodic. Thus, the question concerning the orbital
period of EF~Tuc remains open.

\section{Lib 3 = Preston 874124}
\label{Lib 3}

Lib 3 is quoted as a novalike variable in the catalog of Downes et al.\ (2005).
This classification is based on a private communication of W. Krzeminski cited
by Vogt (1989). While this is compatible with the optical colours 
($B-V=0.15$; $U-B = -0.91$; Beers et al. 1992) the only published spectrum,
showing a blue continuum and weak and broad Balmer absorption lines
(Liu et al. 1999a) is inconclusive.  

I therefore observed a light curve of Lib 3 extending for about 
5\hochpunkt{h}5. During this period the differential magnitude with
respect to the primary comparison star UCAC4 299-243578 remained constant
within the standard deviation of 0\hochpunkt{m}015 of the individual data
points. This complete absence for any sign of flickering during such a long
time is a strong indication that Lib 3 is not a cataclysmic variable.

Since the finding chart published by Downes et al.\ (2005) is based on a
coordinate match instead of a positive identification of a variable star
a mis-identification cannot be excluded. Therefore, I also measured the
differential magnitude with respect to UCAC4 299-243578 of all stars within 
5 arc minutes around the primary candidate which are bright enough to be 
subjected to photometry, considering the limitations of the available data. 
In no case variability was detected.

\section{Variations on short time scales}
\label{Variations on short time scales}

With the exception of Lib~3 all objects of the present study exhibit rather
strong variations on the time scale of minutes. Most of this is expected to
be random flickering as is typical in CVs. However, it is always worthwhile
to investigate if other than apparently stochastic variations hide beneath
the flickering, as is a characterization of the flickering properties
themselves. Therefore, some basic parameters of the rapid light curve 
modulations are determined here. While it is possible to draw some
immediate conclusions, they also serve as input for are more systematic 
quantitative and comparative investigation of the flickering in many 
cataclysmic variables (Bruch, in preparation).

\subsection{Flickering amplitude}
\label{Flickering amplitude}

One of the basic properties of flickering is its total amplitude. However,
if simply taken to be the difference between maximum and minimum magnitude
in a light curve, it will depend to a certain extent on the length of the
data train: the probability to accidentally observe particularly strong 
flickering flares is larger in longer light curves than in shorter ones. 
On the other hand, long light curves may contain variations not related to
flickering (i.e., orbital humps). 

In order to render a determination of the
flickering amplitude comparable between different light curves of the same
object and between different objects, I only regard that part of the flickering
which occurs on time scales of less than 30\hoch{m} (much less than the
typical length of a light curve). To this end, the original
data are smoothed by applying a Fourier filter which effectively removes
all variations on a time scale below 30\hoch{m}. The difference of the
smoothed and the original version then represents a high pass filtered
light curve which is free from variations on longer time scales. Even so the 
difference between maximum and minimum brightness
may still depend strongly on accidental events in the light curve. Therefore,
in order to obtain a more ``typical'' value I take the FWHM of a Gaussian
adjusted to the distribution of the individual magnitude values, as a 
proxy for the flickering amplitude. For each object of the current study the 
average of the corresponding values determined from all available light
curves is listed together with their standard deviation in 
Table~\ref{flickering parameters}. Since data noise tends to broaden the
distribution, it still depends on the noise level in the data which, however,
is similar in all light curves. Applying the formula provided by
Da Costa (1992) the noise in the differential light curves is in all
cases estimated to be of the same order of magnitude (0\hochpunkt{m}01), 
meaning that the numbers in Table~\ref{flickering parameters} are comparable 
among each other. 

Not surprisingly, the flickering amplitude varies significantly 
from one oject to the other. However, for a given star it remains fairly 
constant over time.

\begin{table*}

\caption{Flickering parameters}
\label{flickering parameters}

\hspace{1ex}

\begin{tabular}{lccccc}
\hline
Object & Amplitude & $\alpha_{\rm ps}$  & $\alpha_{\rm wav}$  & $\Sigma$  & 
Number of \\
Name   & (FWHM)(mag) & (power spectrum) & (wavelet) & (wavelet) & 
light curves    \\
\hline
CZ~Aql   & 0.103 $\pm$ 0.019 & -1.76 $\pm$ 0.35 (0.13) & 
1.99 $\pm$ 0.19              & -1.22 $\pm$ 0.11        & \phantom{2}7 \\
BO~Cet   & 0.071 $\pm$ 0.007 & -2.06 $\pm$ 0.26 (0.12) & 
1.82 $\pm$ 0.12              & -1.47 $\pm$ 0.09        & \phantom{2}5 \\
V380~Oph & 0.094 $\pm$ 0.002 & -1.57 $\pm$ 0.09 (0.05) & 
1.52 $\pm$ 0.12              & -1.31 $\pm$ 0.02        & \phantom{2}3 \\
EF~Tuc   & 0.117 $\pm$ 0.016 & -1.52 $\pm$ 0.36 (0.10) & 
1.56 $\pm$ 0.30              & -1.11 $\pm$ 0.11        & 12 \\
\hline
\end{tabular}
\end{table*}
%

\subsection{Red noise behaviour}
\label{Red noise behaviour}

\begin{figure}
   \parbox[]{0.1cm}{\epsfxsize=14cm\epsfbox{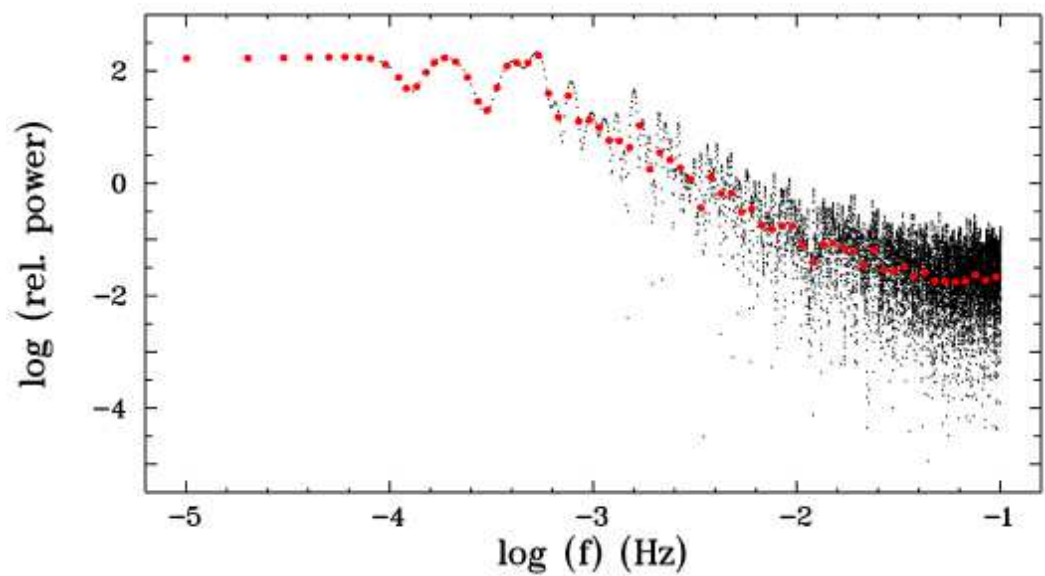}}
      \caption[]{Lomb-Scargle power spectrum of the light curve of BO~Cet
                 observed on 2016, Aug.\ 11 on a double logarithmic scale
                 (black points). The red dots show the same data, binned
                 into frequency intervals of $\Delta \log(f) = 0.05$.}
\label{bocet-power}
\end{figure}

On the double logarithmic scale the power spectra of CV light curves exhibit a
characteristic shape (see e.g. Bruch 1992). It is more or less 
constant at low frequencies (somewhat modulated due to stochastic variations
occurring on the corresponding long time scales). It then declines linearly 
to higher frequencies reflecting red noise behaviour and finally levels off 
to an approximately constant level at very high frequencies due to the 
dominance of white noise at these scales. As an example, in 
Fig.~\ref{bocet-power} the power spectrum of the light curve of BO~Cet of 2016,
Aug.\ 11, is shown, calculated up to the Nyquist frequency, at the original
sampling frequencies (black points) and binned into frequency intervals of
$\Delta \log(f) = 0.05$ (red dots).

An important flickering parameter is the slope $\alpha_{\rm ps}$\footnote{Here,
the subscript ``ps'' is used in order to distinguish this parameter, derived
from a power spectrum analysis, from the scalegram parameter $\alpha_{\rm wav}$,
introduced in Sect.~\ref{Wavelet analysis}, which is based on wavelets.}
of the linear part of 
the power spectrum. It basically measures the distribution of power among
different time scales: The larger $|\alpha_{\rm ps}|$ the more slow flickering 
flares dominate over rapid flares, while small values of $|\alpha_{\rm ps}|$ 
indicate that the amplitudes of slow and rapid flares are more equally 
distributed. 

It is not obvious which is the best way to reliably measure $\alpha_{\rm ps}$ 
such that a comparison between the results for different light curves and 
objects is feasible without depending too strongly on the detailed observing
parameters and conditions. An in-depth investigation of this problem is beyond 
the scope of the present paper and is postponed to a future study. Here, I 
adopt the following procedure: First, all double logarithmic power spectra were 
rebinned, adopting the above mentioned bin width. Thus, in the logarithmic 
representation all bins have the same weight when fitting a function to the 
data. Otherwise, due to the original sampling at constant intervals in 
frequency, high frequencies would get excessive weight. However, since fewer 
data points contribute to the low frequency bins, accidental fluctuations -- 
averaged out in the high frequency bins due to the large number of contributing
data points -- are apt to introduce significant noise even in the
range of $\log f$ where the linear decline has already started. Therefore,
this range has to be avoided when measuring $\alpha_{\rm ps}$. At the other 
extreme it is not trivial to define the frequency at which the power spectrum 
turns flat due to white noise. A visual inspection of all power spectra showed 
that the frequency range $-3 < \log[f (Hz)] < -2$ is least influenced by either 
effect and is thus best suited for a linear fit\footnote{This holds true for the
present data but may be different if light curves observed with other 
instruments and under different conditions are regarded!}.

Table~\ref{flickering parameters} lists for each of the target stars the 
average value for $\alpha_{\rm ps}$ measured in this way. The standard 
deviations $\sigma_{\alpha_{\rm ps}}$
calculated from results for the individual light curves are quite large.
In order to investigate if this is due to real night-to-night variations,
or if the random distribution of flickering flares in a given light curve
can lead to large variations of the measured $\alpha_{\rm ps}$, for 12 of the 
longest light curves $\alpha_{\rm ps}$ was determined separately for the first 
and the second half. The standard deviation of the difference with respect 
to $\alpha_{\rm ps}$ calculated from the entire data set amounts to 0.26 which 
is of the same order of magnitude as $\sigma_{\alpha_{\rm ps}}$. Thus, 
$\sigma_{\alpha_{\rm ps}}$ appears to reflect the intrinsic uncertainty of 
$\alpha_{\rm ps}$ as opposed to real night-to-night variations.

In order to assess if the average values of $\alpha_{\rm ps}$ for the four CVs 
regarded here differ systematically, it is appropriate to compare the 
difference between the averages 
to the standard error of the mean (i.e., $\sigma_{\alpha_{\rm ps}} / \sqrt{n}$, 
where $n$ is the number of light curves), quoted in brackets in 
Table~\ref{flickering parameters}. It is then seen that the average 
$\alpha_{\rm ps}$ can indeed vary significantly from one object to another. 

\subsection{Wavelet analysis}
\label{Wavelet analysis}

Fritz \& Bruch (1998) pioneered the application of wavelet transforms to 
flickering light curves of many CVs of different subtypes. Later, 
Tamburini et al.\ (2009) extended this work to the intermediate polar
V709~Cas, and Anzolin et al.\ (2010) used similar techniques to investigate
flickering in x-rays of a large sample of CVs.

Fritz \& Bruch (1998) found that the
(logarithmic) scalegram (Scargle et al. 1993) is always largely linear. This 
permits a parameterization in terms of its inclination $\alpha_{\rm wav}$ and 
its value $\Sigma$ (flickering strength) at a reference time scale 
(for details see Fritz \& Bruch 1998). It is found that the location of an 
object in the $\alpha_{\rm wav} - \Sigma$ plane depends on the CV subtype and 
its photometric state. 

\begin{figure}
   \parbox[]{0.1cm}{\epsfxsize=14cm\epsfbox{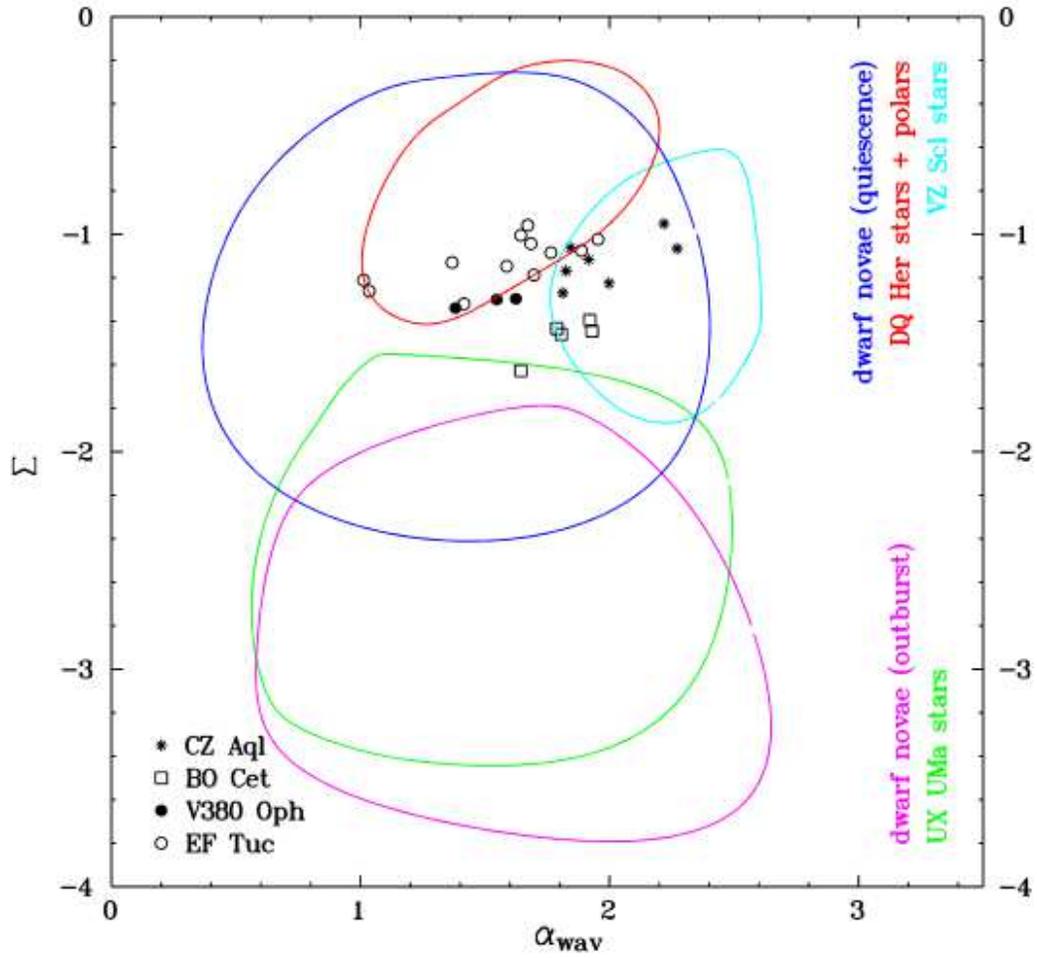}}
      \caption[]{Scalegram parameters $\alpha_{\rm wav}$ vs. $\Sigma$ derived
                 from a wavelet analysis of the light curves of the objects
                 of this study. The coloured contours indicate the limits
                 in the $\alpha_{\rm wav} -\Sigma$ plane of the range populated
                 by CVs of different kinds and photometric states as found by
                 Fritz \& Bruch (1998).}
\label{wavelet}
\end{figure}

Applying the same procedures adopted by Fritz \& Bruch (1998) to the present 
light curves it is found that $\alpha_{\rm wav}$ and $\Sigma$ are fairly stable
over time with average values as quoted in Table~\ref{flickering parameters}.
The values derived from the individual light curves are shown in
Fig.~\ref{wavelet}. Since Fritz \& Bruch (1998) showed that the location
of CVs in the $\alpha_{\rm wav} - \Sigma$ plane depends on their subtypes
and photometric states the limits of the ranges populated by different CVs 
populations are also shown in the figure (adapted from Figs. 10 -- 14 of 
Fritz \& Bruch 1998). This provides some clues to the nature of the 
investigated systems. 

All objects lie comfortably in the (wide) range occupied by quiescent
dwarf novae. However, this does not permit a unique classification and
a closer inspection is in order.

Not much is known about the long term brightness variations of CZ~Aql. The
AAVSO data base contains only a few data points scattered within a magnitude
range compatible with the observed flickering. No outbursts were ever reported.
Based on spectroscopic evidence, Sheets et al.\ (2007) suspected a 
magnetically channeled accretion flow. In Fig.~\ref{wavelet}
CZ~Aql lies just outside the range of DQ~Her stars (intermediate polars) and
polars. It is, however, within the range of VZ~Scl novalike variables. Long 
term monitoring in order to detect either outbursts or low states should 
permit to better determine the subtype of CZ~Aql.

With a single outlying light curve, BO~Cet lies in the range of VZ~Scl 
novalike variables. At first sight this appears to be slightly at odds with its
classification as SW~Sex star (Rodr\'{\i}guez-Gil et al. 2007), all such 
systems included in the study of Fritz \& Bruch (1998) having significantly 
smaller values of $\Sigma$. However, as Rodr\'{\i}guez-Gil et al.\ (2007) 
pointed out, half of the known VY~Scl systems are also SW~Sex stars, but none 
of the SW~Sex stars in the sample of Fritz \& Bruch (1998) is known to exhibit 
low states. Again, the AAVSO data base is insufficient to draw any conclusion, 
while a light curve generated from AEOEV data exhibits variability between 
$\sim$13\hochpunkt{m}7 and $\sim$15\hochpunkt{m}2 but is not dense enough
to permit the identification of outbursts or low states.

The long term photometric as well as the spectroscopic evidence cited in
Sect.~\ref{V380 Oph} clearly identify V380~Oph as both, a SW~UMa and VZ~Scl
type star. In the $\alpha_{\rm wav} - \Sigma$ plane it lies somewhat beyond
the VZ~Scl borders, which might just mean that the sample of 
Fritz \& Bruch (1998) is too limited to encompass the entire range.

Finally, as shown in Sect.~\ref{EF Tuc}, EF~Tuc is a genuine dwarf nova and
this is corrobrated by its position in the $\alpha_{\rm wav} - \Sigma$ plane.

\subsection{The intermediate and high frequency regime}
\label{The intermediate and high frequency regime}

Looking for consistent variations on time scales as expected to be
caused, e.g., by the rotation of the white dwarf in intermediate polars or
by Quasi Periodic Oscillations (QPOs) it may be misleading to turn to the
power spectrum of an entire light curve. Random flickering flares,
eventually accidentally recurring a few times quasi-periodically within a 
limited time interval, can easily cause strong power spectrum peaks which
may be mistaken as indications for persistent oscillations. It is
therefore more appropriate to regard time resolved power spectra. 

\begin{figure}
   \parbox[]{0.1cm}{\epsfxsize=14cm\epsfbox{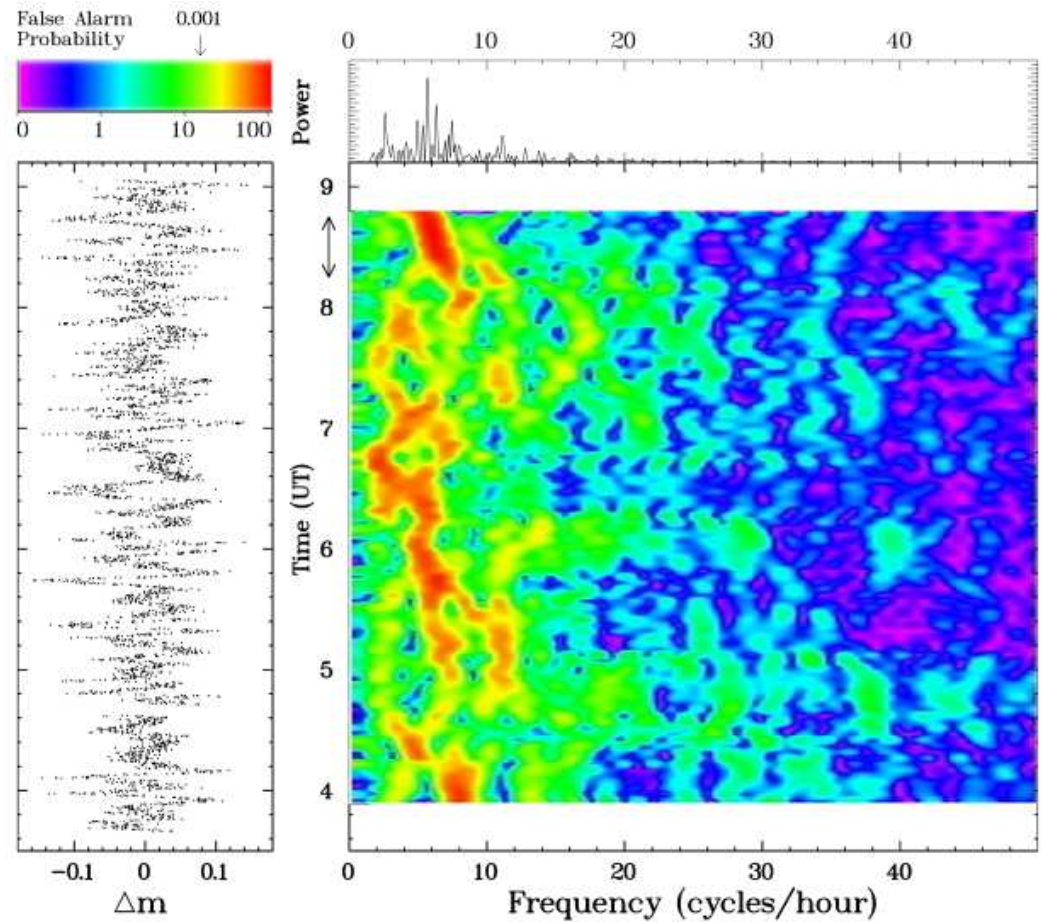}}
      \caption[]{Light curve of CZ~Aql of 2014, June 17 after removal of
                variations on time scales above 30\hoch{m} (left frame) and 
                stacked power spectra (relative power on a logarithmic scale)
                of the same
                data (right). Spectral features within a range of 30\hoch{m} 
                are not independent from each other. The length of this range
                is indicated by a double arrow at the upper left margin of the
                stacked power spectra. The upper right frame contains the
                power spectrum calculated from the entire data set. At the
                upper left, a color bar is shown, where the false alarm
                probability level of 0.001 is marked.}
\label{czaql-power}
\end{figure}

To this end, stacked power spectra was calculated
as outlined by Bruch (2014): Using the high pass filtered light
curves introduced in Sect.~\ref{Flickering amplitude},
sections of $\Delta \tau = 30\hoch{m}$ duration and an overlap of 27\hoch{m} 
between successive sections were taken. For each of them a
Lomb-Scargle periodogram was calculated and the individual power spectra 
were stacked on top of each other to result in a two-dimensional 
representation (frequency vs.\ time). An example is shown in 
Fig.~\ref{czaql-power} which refers to the light curve of CZ~Aql of 2014,
June 17. Relative power is shown on a logarithmic colour scale. To the left 
of the stacked power spectrum the high pass filtered light curve is shown,
and above it the Lomb-Scargle periodogram of the entire data set is plotted. 
Due to the chosen values for
$\Delta \tau$ structures on time scales of less than 30\hoch{m} (indicated by 
the double-headed arrow at the upper left of the time axis) are not independent.
The false alarm probability of 0.001 for power spectrum peaks is marked 
on the colour bar in the upper left frame of the figure. It was calculated
using Eq.~18 of Scargle (1982), the number of independent frequencies
having been determined as described by Bruch \& Diaz (2917). Thus, the yellow
and red structures are highly significant.

In none of the stacked power spectra features were observed which might 
confidently be interpreted as QPOs. However, while not conclusive, the results
for CZ~Aql may point at an intermediate polar nature as implied by the
suspicion of Sheets et al.\ (2007) of a magnetically channeled accretion
flow in this system. In all of the stacked power spectra strong structures
occur preferentially at frequencies corresponding to periods between 
9\hochpunkt{m}5 and 12\hochpunkt{m}2 which often extend in time for
intervals $\gg \Delta \tau$. An example can be seen in the central part 
of the stacked power spectrum shown in Fig.~\ref{czaql-power}. A strong
signal with at a similar frequency again appears at the end of the light
curve. Together, they lead to the main peak in the Lomb-Scargle periodogram
in the upper frame of the figure, which corresponds to a period of 
10\hochpunkt{m}6. The varying strength of such features and their shifts in 
frequency in the stacked power spectra can be explained
by the influence of random variations: Tests with artificial sinusoidal
signals even with large amplitudes (up to 50\% of the total flickering
amplitude) added to the real data showed that their reflection in the stacked
spectra can easily be suppressed or frequency shifted during some time interval 
by the influence of flickering.  

\begin{figure}
   \parbox[]{0.1cm}{\epsfxsize=14cm\epsfbox{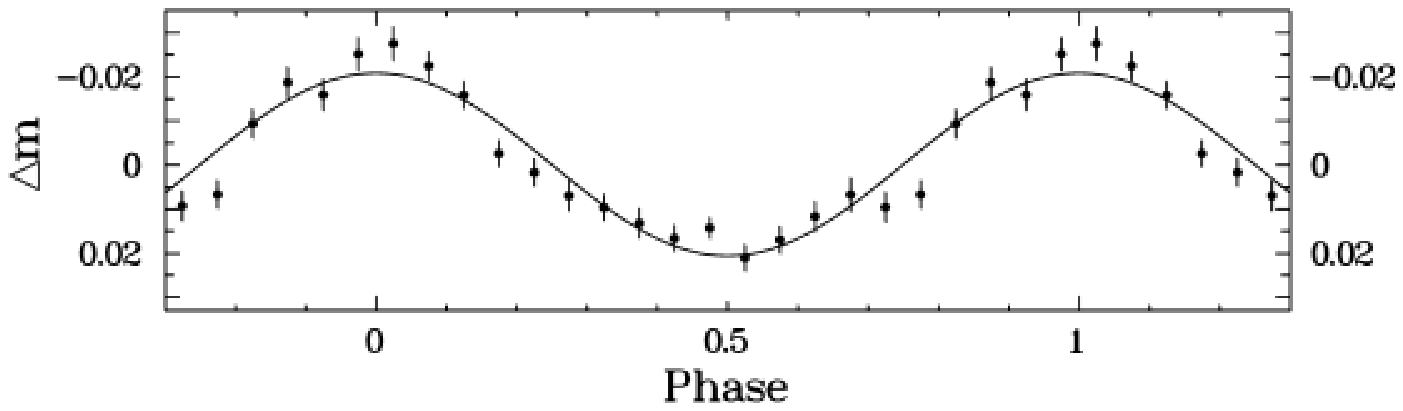}}
      \caption[]{Light curve of CZ~Aql of 2014, June 17 (dots) after removal
                of variations on time scales above 30\hoch{m}, folded on
                period 10\hochpunkt{m}6 and binned in intervals of width 
                0.05 in phase. The error bars represent mean errors of the mean.
                The solid line is a least squares sine fit the the data.}
\label{czaql-fold}
\end{figure}

In Fig.~\ref{czaql-fold} the high pass filtered light curve of CZ~Aql of
2014, June 17, folded on the suggested period of 10\hochpunkt{m}6 and
binned into phase intervals of 0.05, is shown together with the best fit sine 
curve. Interpreting the variations as due to the rotation of the white dwarf,
the spin to orbital period ratio is thus $P_{\rm spin}/P_{\rm orb}
\approx 0.037$ which is totally in line with ratio observed for the
majority of intermediate polars (see, e.g., Koji Mukai's Intermediate
Polar Homepage\footnote{asd.gsfc.nasa.gov/Koji.Mukai/iphome/iphome.html}).
The total amplitude of 0\hochpunkt{m}04 is also not unusual for IPs with
similar spin period
(MU~Cam, Staude et al. 2003; PQ~Gem, Evans et al. 2006; NY~Lup, Haberl et al.
2002). 

While by far not sufficient to classify CZ~Aql as an intermediate polar, 
the evidence presented here, together with the proximity of the object to
the intermediate polars in the $\alpha_{\rm wav} - \Sigma$ plane (see
Sect.~\ref{Wavelet analysis}), the spectroscopic indications quoted by 
Sheets et al.\ (2007) and the presence of a rather strong He~II 
$\lambda$ 4686 \AA\, emission line seen in the spectrum of
Cieslinski et al.\ (1998) -- a feature not exclusive to but typical for
magnetic CVs -- justifies to
further consider this hypothesis. To prove or disprove it will require
additional observations.   

\section{Summary}
\label{Summary}

I have presented the first time resolved photometry for an ensemble of
five cataclysmic variables and candidates. While four of them (CZ~Aql,
BO~Cet, V380~Oph and EF~Tuc) are established members of the class, based
on previous spectroscopic studies, Lib~3 was only suspected to be a CV. 
The absence of flickering -- a {\em sine qua non} for all CVs -- in the 
latter star justifies to remove Lib~3 from the list of CV candidates.

All other targets of this study exhibit pronounced flickering. This is
an indication that their accretion disks are not in a stable bright
state as is observed in nova-like variables of the UX~UMa class or most
old novae, which have much smaller flickering amplitudes (Fritz \& Bruch 1998).
The flickering behaviour rather points at a VY~Scl type classification
for BO~Cet and V380~Oph (in the latter case confirmed by its long-term
light curve), in addition to their classification as SW~Sex stars
(Rodr\'{\i}guez-Gil et al.\ (2007). EF~Tuc is a genuine dwarf nova, observed in 
quiescence, while the evidence for CZ~Aql is unclear. It may be a VY~Scl type 
star or a dwarf nova. Additionally, a nature as intermediate polar cannot be 
discarted.

Variations on hourly time scales reveal clear orbital modulations in BO~Cet
during the 2014 observing season, which, however, are not as clear in 2016.
Orbital modulations were probably also observed in V380~Oph. On the other
hand, a modulation with a period $\sim$8\% longer than the orbital period
was observed in CZ~Aql which may be similar to variations interpreted as
superhumps in several other (non SU~UMa type) system, many of which are known
VY~Scl stars.    

\section*{Acknowledgements}

I gratefully acknowledge the use of the AAVSO, AFOEV and BAAVSS data bases 
which provided valuable supportive information for this study.

\section*{References}

\begin{description}

\item
      Allen, C.W. 1973, Astrophysical Quantities, third edition 
      (Athlone Press: London)
\item
      Anzolin, G., Tamburini, F., de Martino, D., \& Bianchini, A. 2010,
      A\&A 519, A69
\item
      Beers, T.C., Preston, G.W., Shectman, S.A., Doinidis, S.P., \&
      Griffin, K.E. 1992, AJ, 103, 267
\item
      Belova, A.I., Suleimanov, V.F., Bikmaev, I.F., et al. 2013, Astron.\ 
      Letters, 39, 111
\item
      Bruch, A. 1991, Acta Astron, 41, 101 
\item
      Bruch, A. 1992, A\&A 266, 237
\item
      Bruch, A. 1993, 
      A Reference Guide (Astron.\ Inst.\ Univ.\ M\"unster
\item
      Bruch, A. 2014, A\&A, 566, A101
\item
      Bruch, A. 2016, New Astr., 46, 90
\item
      Bruch, A., Diaz, M.P. 2017, New Astr., 50, 109
\item
      Chen, A., O'Donogue, D., Stobie, R.S., Kilkenny, D., Warner, B. 2001
      MNRAS, 385, 89
\item
      Cieslinski, D., Steiner, J.E., \& Jablonski, F.J. 1998, A\&AS 131, 119
\item
      Da Costa, G.M. 1992, ASP Conf.\ Ser., 23, 90
\item
      Downes, R.A., Webbink, R.F., Shara, M.M., et al. 
      2005, J.\ Astron.\ Data, 11, 2
\item
      Eastman, J., Siverd, R., \& Gaudi, B.S. 2010, PASP, 122, 935
\item
      Evans, P.A., Hellier, C., \& Ramsay, G. 2006, MNRAS, 369, 1229
\item
      Fritz, T., Bruch, A. 1998, A\&A, 332, 586
\item
      Garcia, A, Sodr\'e Jr., L., Jablonski, F.J., \&
      Terlevich, R.J. 1999, MNRAS, 309, 803
\item
      Gromadzki, M., Miko{\l}ajewski, M., Tomov, T., et al. 2006,
      Acta Astron., 56, 97
\item
      Haberl, F., Motch, C., \& Zickegraf, F.-J. 2002, A\&A, 387, 201
\item
      Haefner, R., Metz, K. 1985, A\&A, 145, 311
\item
      Herbst, W., \& Shevchenko, V.S. 1999, AJ, 118, 1043
\item
      Hoffmeister, C. 1929, Mitt.\ Sternw.\ Sonneberg, N16
\item
      Horne, J.H., \& Baliunas, S.L. 1986, ApJ, 302, 757
\item
      Kafka, S., \& Honeycutt, R.K. 2004, Rev.\ Mex.\ A\&A (Conf.\ Series),
      20, 238
\item
      Kato, T., Imada, A., \& Uemura, M. 2009, PASJ, 61, S395
\item
      Kazarovetz, E.V., Samus, N.N., \& Goranskij, V.P. 1993, IBVS 3840
\item
        Kenyon, S.J., Kolotilov, E.A., Ibragimov, M.A., \&
        Mattei, J.A. 2000, ApJ, 531, 1028
\item
      Kozhevnikov, V.P. 2007, MNRAS, 378, 955
\item
      Kozhevnikov, V.P. 2012, New Astron., 17, 38
\item
      Liu, W., Hu, J.Y., Li, Z.Y., \& Cao, L. 1999a, ApJS, 122, 257
\item
      Liu, W., Hu, J.Y., Zhu, X.H., \& Li, Z.Y. 1999b, ApJS, 122, 243
\item
      Lomb, N.R. 1976, Ap\&SS, 39, 447
\item
      Papadaki, C., Boffin, H.M.J., Stanishev, V., et al.\ 2009, 
      J.\ Astron.\ Data, 15, 1
\item
      Papadaki, C., Boffin, H.M.J., Sterken C., et al.\ 2006, A\&A, 456, 599
\item
      Patterson, J., Thomas, G., Skillman, D.R., Diaz, M. 1993, ApJ Suppl., 
      86, 235
\item
      Patterson, J., Thorstensen J.R., Fried, R., et al. 2001, PASP, 113, 72 
\item
      Reinmuth, K. 1925, Astron,\ Nachr.\ 225, 385
\item
      Ritter, H., Kolb, U. 2003, A\&A, 404, 301
\item
      Rodr\'{\i}guez-Gil, P., Schmidtobreik, L., \& G\"ansicke, B. 2007
      MNRAS, 374, 1359
\item
      Scargle, J.D. 1982, ApJ, 263, 853
\item
      Scargle, J.D., Steiman-Cameron, T.Y., Young, K., et al. 1993, 
      ApJ 411, L91
\item
      Scaringi, S., Maccarone, T.J., K\"ording, E., et al. 2015, 
      Science Adv., 1, e1500686
\item
      Schwarzenberg-Czerny, A. 1989, MNRAS 241, 153
\item
      Shafter, A.W. 1983, IBVS, 2377
\item
      Shafter, A.W. 1985, AJ, 90, 643
\item
      Sheets, H.A., Thorstensen, J.R., Peters, C.J., \& Kapusta, A.B.
      2007, PASP, 119, 494
\item
      Shugarov, S.Yu., Katysheva, N.A., Seregina, T.M., \& Volkov,I.M.
      2005, in: The Astrophysics of Cataclysmic Variables and Related Objects
      (eds: J.-M.\ Hameury \& J.-P.\ Lasota), ASP Conf.\ Series, 330, p.\ 495
\item
      Smak, J. 2013, Acta Astron., 17, 453
\item
      Staude, A, Schwope, A.D., Krumpe, M., Hambaryan, V., \& Schwarz, R.
      2003, A\&A, 406, 253
\item
      Stellingwerf, R.F. 1978, ApJ, 224, 953
\item
      Stobie, R.S., Kilkenny, D., O'Donoghue, D. 1995, ApSS, 230, 101
\item
      Tamburini, F., de Martino, D., \& Bianchini, A. 2010, A\&A 502, 1
\item
      van der Klis, M. 2004, arXiv e-prints {\tt[arXiv:astro-ph/0410551]}
\item
      Verbunt, F., Bunk, W.H., Ritter, H., \& Pfeffermann, E. 1997,
      A\&A 327, 602
\item
      Vogt, N. 1989, in: Classical Novae, ed. M.F.\ Bode and A.\ Evans 
      (New York, Wiley and Sons), p.\ 225
\item
      Zacharias, N., Finch, C.T., Girard, T.M., et al.\ 2013, AJ, 145, 44
\item
      Zwitter, T., \& Munari, U. 1995, A\&AS, 114, 575

\end{description}

\end{document}